\documentstyle[prl,aps,epsf,twocolumn]{revtex}
\def\eq{\begin{equation}}
\def\be{\begin{equation}}
\def\ee{\end{equation}}
\def\eqa{\begin{eqnarray}}
\def\eea{\end{eqnarray}}
\def\ra{\rightarrow}

\def\s{\sigma}
\def\p{\partial}

\def\w{\omega}

\begin{document}
\draft
\flushbottom
\twocolumn[
\hsize\textwidth\columnwidth\hsize\csname @twocolumnfalse\endcsname
\title{Low Energy Properties of the (n,n) Carbon Nanotubes}
\author{Yu.~A.~Krotov $^{(a)}$, D.-H.~Lee $^{(a)}$, and Steven~G.~Louie $^{(a,b)}$} 
\address{$^{(a)}$
Department of Physics, University of California at Berkeley, Berkeley, CA 94720}
\address{$^{(b)}$
Materials Sciences Division, Lawrence Berkeley National Laboratory, Berkeley, CA 94720}
\date{\today}
\maketitle
\tightenlines
\widetext
\advance\leftskip by 57pt
\advance\rightskip by 57pt

\begin{abstract}
According to band theory, an ideal undoped (n,n) carbon nanotube is metallic. We show that the electron-electron interaction causes it to become Mott insulating with a spin gap. More interestingly, upon doping it develops superconducting fluctuations. 
\end{abstract}

\vskip 1cm

]

\narrowtext
\tightenlines

Carbon nano-structures such as $C_{60}$ \cite{c60} and nanotubes \cite{disc}  have attracted considerable interest recently. The latter is a graphite sheet wrapped into a cylinder form. A a pair of integers $(n,m)$ specifies the wrapping. Starting from a graphite sheet with the primitive lattice vectors $\vec{a},\vec{b}$ making an angle of 60 degrees, the $(n,m)$ tube is a cylinder with axis running  perpendicular to $n\vec{a}+m\vec{b}$, so that atoms separated by $n\vec{a}+m\vec{b}$ are wrapped onto each other. 

Considerable efforts have gone into studying the bandstructure of carbon nanotubes\cite{ab}. The purpose of this work is to address the effects of electron-electron interaction on the low energy properties of them. In one-electron tight-binding description where one retains a single $\pi$ orbital per atom and keeps only the nearest neighbor hopping, all $n\ne m$ tubes are band-insulators with gaps generally scaling inversely with the radius\cite{ab}. For these tubes a sufficiently weak electron-electron interaction is not expected to change the low energy properties qualitatively. The same can not be said of the (n,n) tubes, which are band metals. To be precise, for the latter two out of $4n$ bands intersect the Fermi level to form two Dirac points (Fig.1). Due to the low dimensionality and the presence of gapless excitations, the effects of electron-electron interaction must be examined more carefully.

\begin{figure}[b]
\epsfysize=4.5cm\centerline{\epsfbox{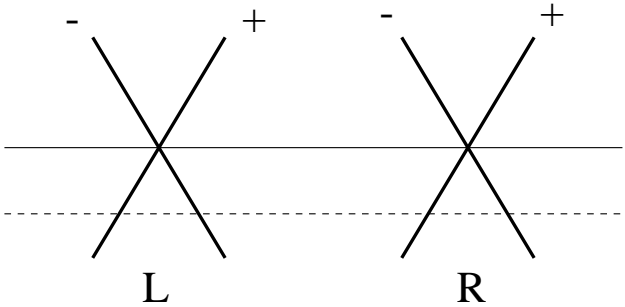}}
\vspace{20pt}
\caption{ The low-energy band structures of (n,n) tubes. ``L'' and ``R'' labels the Dirac points, ``+'', ``-'' labels the right and left movers. The dashed line denotes the Fermi level in the doped case.}
\label{fig1}
\end{figure}

In this paper we perform perturbative renormalization group (RG) calculations to analyze the asymptotic low energy behavior of the (n,n) tubes\cite{exp}. The microscopic Hamiltonian is consisted of a nearest-neighbor tight-binding model for the $\pi$ orbital on each carbon atom (the bonding topology is illustrated in Fig. 2) and a Hubbard $U$ plus a nearest neighbor $V$ for electron correlations. To build up an effective model for low temperature, we first discard all bands that do not intersect the Fermi level. Second, we  regard the two remaining bands as linearly dispersing,  i.e., we ignore the energy dependence of the band velocity. Due to these approximations, a upper energy cutoff $E_c$ has to be imposed on our subsequent discussions \cite{note}.  In the Hilbert space of the two bands, the original interactions U and V give rise to twelve independent scattering amplitudes $g^{abcd}_{ijkl}$.

\begin{figure}[b]
\epsfysize=4cm\centerline{\epsfbox{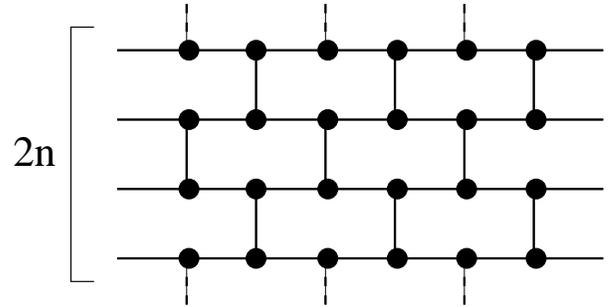}}
\vspace{20pt}
\caption{ The bonding structure of (n,n) tubes.}
\label{fig2}
\end{figure}

The effective Hamiltonian is given by $H=H_K+H_I$, where
\eqa
H_K&=&\sum_{i=R,L}\sum_{a=\pm}\sum_{\sigma}(av_F)\int dx\psi_{ia\sigma}^{\dagger}(x)\frac{\p_x}{i} \psi_{ia\sigma}(x)\nonumber \\
H_I&=&\sum\frac{
g^{abcd}_{ijkl}}{2} \int dx\psi_{ia\sigma}^{\dagger}(x) \psi_{jb\sigma'}^{\dagger}(x) \psi_{kc\sigma'}(x) \psi_{ld\sigma}(x).
\label{h}
\eea
In the second line of Eq.(\ref{h}) the sum is taken over  $i,j,k,l=R,L$, $a,b,c,d=\pm$, and $\s,\s'=\uparrow,\downarrow$. The operator 
$\psi^+_{i\pm\sigma}(k)=\int dx e^{ikx}\psi_{i\pm\sigma}^{\dagger}(x)$ creates an right/left moving electron with momentum $k_i+k$ and with spin $\s$. Here $k_i$ is the momenta associated with the right (R) or left (L) Dirac points (Fig.1).
The only nonzero $g$'s in Eq.(\ref{h}) are those whose lower and upper indices have the form: $(RRRR),(RLRL),(RLLR)$, and 
$(++++),(++--),(+-+-),(+--+)$, plus those generated by $+\leftrightarrow -$ and/or $R \leftrightarrow L$. We abbreviate the twelve independent coupling constant as $g_i^j$, where $i=1,2,4$; $j=1,2,3,4$ \cite{note2}.

We then perform a one-loop RG calculation, which is a straightforward generalization of the one-band calculation \cite{Shankar}.  The only complication is that instead of four, there are twelve independent scattering amplitudes. This calculation yields:   
\eqa
&&\frac{dg_1^1}{dx} = -g_1^3g_1^3 + g_1^3g_2^3 -g_1^1g_1^1 - g_1^2g_2^1, 
\nonumber \\
&&\frac{dg_1^2}{dx} = g_4^2g_1^2 + g_4^3g_1^3 - g_1^2g_2^2 - g_2^1g_1^1, \nonumber \\
&&\frac{dg_1^3}{dx} = -2g_1^3g_1^1 + g_1^3g_2^2 + g_2^3g_1^1 + g_4^2g_1^3 + 
g_4^3g_1^2, \nonumber \\
&&\frac{dg_2^1}{dx} = g_4^2g_2^1 + g_4^1g_1^2 - 2g_4^1g_2^1 + g_4^3g_1^3 - 
g_4^3g_2^3 - g_1^2g_1^1- g_2^1g_2^2, \nonumber \\
&&\frac{dg_2^2}{dx} = \frac{1}{2} (g_2^3g_2^3 - g_1^2g_1^2 - g_2^1g_2^1 - g_1^1g_1^1),\nonumber \\
&&\frac{dg_2^3}{dx} = g_2^3g_2^2 + g_4^2g_2^3 + g_4^1g_1^3 - 2g_4^1g_2^3 + 
g_4^3g_1^2 - g_4^3g_2^1, \nonumber \\
&&\frac{dg_4^1}{dx} = -g_4^1g_4^1 + g_1^3g_2^3 -g_2^3g_2^3 + g_1^2g_2^1 -g_2^1g_2^1, \nonumber\\
&&\frac{dg_4^2}{dx} = \frac{1}{2} (g_4^3g_4^3 - g_4^1g_4^1 + g_1^3g_1^3 + g_1^2g_1^2), \nonumber \\
&&\frac{dg_4^3}{dx} = -g_4^3g_4^1 + 2g_4^3g_4^2 - 2g_2^3g_2^1 + g_1^3g_2^1 + 
g_2^3g_1^2 + g_1^3g_1^2,
\label{rg}
\eea
plus $\frac{dg_1^4}{dx}=\frac{dg_2^4}{dx}=\frac{dg_4^4}{dx} =0$.
In the above $x\equiv\frac{1}{\pi v_F}ln(\frac{E_c}{E})$, where $E_c$ and $E$ are initial and running energy cutoffs respectively.

To draw the implications from the RG equations 
we compute correlation functions  $\chi_{\alpha}(\w)=\int dxdt e^{i\w t}<T[O_{\alpha}(x,t)O^+_{\alpha}(0,0)]>$ for the following $O_{\alpha}$:
\eqa
&&O_{CDW1}(x)=\sum_{\s}\psi_{R+\sigma}^\dagger(x)\psi_{R-\sigma}(x), \nonumber \\
&&O_{SDW1}(x)=\frac{1}{2}[\psi_{R+\uparrow}^\dagger(x)\psi_{R-\uparrow}(x)-\psi_{R+\downarrow}^\dagger(x)\psi_{R-\downarrow}(x)]\nonumber \\
&&O_{SS1}(x)= \frac{1}{\sqrt{2}}[\psi_{R+\uparrow}^\dagger(x) \psi_{R-\downarrow}^\dagger(x) -\psi_{R+ \downarrow}^\dagger(x) \psi_{R-\uparrow}^\dagger(x) ] , \nonumber \\ 
&&O_{TS1}(x)=\psi_{R+\uparrow}^\dagger(x)\psi_{R-\uparrow}^\dagger(x), \nonumber \\
&&O_{CDW2}(x)=\sum_{\s}\psi_{R+\sigma}^\dagger(x)\psi_{L-\sigma}(x), \nonumber \\
&&O_{SDW2}(x)=\frac{1}{2}[\psi_{R+\uparrow}^\dagger(x)\psi_{L-\uparrow}(x)-\psi_{R+\downarrow}^\dagger(x)\psi_{L-\downarrow}(x)], \nonumber \\
&&O_{SS2}(x)= \frac{1}{\sqrt{2}}[\psi_{R+\uparrow}^\dagger(x) \psi_{L-\downarrow}^\dagger(x)- \psi_{R+\downarrow}^\dagger(x)\psi_{L-\uparrow}^\dagger(x) ] , \nonumber \\ 
&&O_{TS2}(x)= \psi_{R+\uparrow}^\dagger(x)\psi_{L-\uparrow}^\dagger(x).
\eea
plus those generated by $R\leftrightarrow L$ and/or $+\leftrightarrow -$. In the above CDW, SDW, SS and TS stand for charge density wave, spin density wave, singlet superconductivity and triplet superconductivity respectively. These correlation functions are chosen because of their logarithmic divergence in the absence of interaction. 
To the lowest order in $g$'s, the results are
\eq
\chi_{\alpha}=N_{\alpha}\chi_0(\omega)(\frac{E_c}{\omega})^{K_{\alpha}}.
\ee
Here $\chi_0=\frac{1}{2\pi v_F}\ln(\frac{E_c}{\omega})$, and all $N_{\alpha}$ are unity except that $N_{CDW1}=N_{CDW2}=2$.
Finally the exponents are 
$K_{CDW1}=\frac{g_4^2-2g_4^1}{2\pi v_F}, K_{SDW1}=\frac{ g_4^2}{2\pi v_F}, K_{SS1}=-\frac{g_4^2+g_4^1}{2\pi v_F},
K_{TS1}=\frac{g_4^1-g_4^2}{2\pi v_F}, K_{CDW2}=\frac{g_2^2-2g_1^1}{2\pi v_F}, K_{SDW2}=\frac{g_2^2}{2\pi v_F},
K_{SS2}=-\frac{ g_2^2+g_1^1}{2\pi v_F}, K_{TS2}=\frac{ g_1^1-g_2^2}{2\pi v_F}$.

{\bf{I. The undoped case.}}
\rm
The bare values of the coupling constants are 
$ g_2^1=g_2^3=g_4^1=g_4^3=U-V$, $ g_1^1= g_1^2=g_1^3=g_1^4=U$, $g_2^2=g_2^4=g_4^2=g_4^4=U+V$.
We solve Eq.(\ref{rg}) numerically for $V/U\le 1$ and find that in all cases the absolute values of all coupling constants eventually diverge. As to the susceptibilities, 
we find that for $\frac{V}{U}<(\frac{V}{U})_c\approx 0.8$ the most divergent susceptibility is $\chi_{SDW1}$, and for   $(\frac{V}{U})_c<\frac{V}{U}<1$ the most divergent susceptibility is $\chi_{CDW2}$.

{\bf {Ia)The SDW1 Phase:}}
\rm
For $\frac{V}{U}<(\frac{V}{U})_c$ we perform a mean-field theory using the {\it renormalized} Hamiltonian. Guided by the susceptibility result, we introduce $M_R=<O_R(x)>$ and $M_L=<O_L(x)>$, where
\eqa
&&O_R(x)=\frac{1}{2}[\psi_{R+\uparrow}^+(x)\psi_{R-\uparrow}(x)- \psi_{R+\downarrow}^+(x)\psi_{R-\downarrow}(x)]\nonumber \\ &&O_L(x)=\frac{1}{2}[\psi_{L+\uparrow}^+(x)\psi_{L-\uparrow}(x)- \psi_{L+\downarrow}^+(x)\psi_{L-\downarrow}(x)]
\label{order}
\eea
as the order parameters. The mean-field Hamiltonian is the factorized form 
\eqa
H&=&H_K- g_4^3\int dx[O_R O_R +O^+_R O^+_R + O_L O_L + O^+_L O^+_L]
\nonumber \\
&-&2g_4^2\int dx[O_R^+ O_R + O_L^+ O_L]-2g_1^3[O_L O_R + O_R^+ O_L^+]\nonumber \\
&-&2g_1^2\int dx[O_L^+ O_R + O_R^+ O_L].
\label{mf1}
\eea
In the range of $V/U$ considered here the renormalized values of all the $g$'s appearing in Eq.(\ref{mf1}) are positive. Consequently, the mean-field solution predicts $M_R=M_R^*=M_L=M_L^*=M_0$. It is important to note that the term proportional to $g_4^3$ demands both $M_R$ and $M_L$ to be real and the term proportional to $g_1^2$ requires them to have the same sign. Consequently, the mean-field theory completely fixes the U(1) phases of the order parameters $M_R$ and $M_L$. The only global degree of freedom left is the SU(2) rotation of the order parameter away from the spin z-direction. The Goldstone mode associated with the latter governs the low energy physics of the SDW1 phase.
Indeed, let us define
\eqa
&&Q_{\s\s'}(x)=
<\psi_{R+\s}^\dagger(x)\psi_{R-\s'}(x) -\frac{\delta_{\s\s'}}{2}\psi_{R+\tau}^\dagger(x)\psi_{R-\tau}(x)>\nonumber \\
&&+<\psi_{L+\s}^\dagger(x)\psi_{L-\s'}(x)-\frac{\delta_{\s\s'}}{2}\psi_{L+\tau}^\dagger(x)\psi_{L-\tau}(x)>+ h.c.
\eea
The repeated indices imply summation. 
The mean field solution corresponds to $Q_{\s\s'}=Q^0_{\s\s'}=2\s M_0\delta_{\s\s'}$. A smooth twist  in the direction of the order parameter corresponds to 
$Q_{\s\s'}=[U^+Q^0U]_{\s\s'}$, where $U$ is a smooth space-time dependent SU(2) matrix.
After a proper rescaling of the space and time coordinates,
the effective action governing the dynamics of such twists is 
\eq
S_{\s}=\frac{g}{2}\int dtdx Tr[(\p_{\mu}Q)^2].
\label{sigma}
\ee
Unlike the half-filled one-band Hubbard model, the
effective action does not contain a topological term.
This is because there are two Dirac points instead of one. Indeed, it can be shown that each Dirac point contributes a topological term with coefficient $\pi$ \cite{nagaosa}. Thus when added together, their effects cancel! \cite{dhl} 
The 1+1 dimensional non-linear $\s$ model given by Eq.(\ref{sigma}) is always disordered \cite{polyakov}, which indicates a gap in the spin spectrum\cite{haldane}.

{\bf{Ib) The CDW2 Phase:}}
\rm
For $1>\frac{V}{U}>(\frac{V}{U})_c$, the most divergent susceptibility is $\chi_{CDW2}$.
Therefore we introduce $\Delta_1=<O_1(x)>$ and $\Delta_2=<O_2(x)>$ where
\eqa
&&O_1(x)=\sum_{\s}\psi_{R+\sigma}^\dagger(x)\psi_{L-\sigma}(x) \nonumber \\
&&O_2(x)=\sum_{\s}\psi_{L+\sigma}^\dagger(x)\psi_{R-\sigma}(x),
\eea
as order parameters.
The mean-field Hamiltonian is the factorized form of 
\eqa
H&&=H_K+\frac{1}{2}(2g_1^3-g_2^3)\int dx[O_1 O_2 +O^+_1 O^+_2]\nonumber \\
&&+\frac{1}{2}(2g_1^1-g_2^2)[O_1^+ O_1 + O^+_2 O_2].
\label{mf2}
\eea
For the range of $V/U$ considered here the renormalized
$(2g_1^3-g_2^3)$ and $(2g_1^1-g_2^2)$ are positive and negative respectively. In the mean-field theory the term proportional to $(2g_1^3-g_2^3)$ sets the sum of the phases of $\Delta_1$ and $\Delta_2$ to $\pi$, i.e., $\Delta_1=-\Delta_2^*$. The mean-field solution is not unique, there being a family of equivalent mean-field solutions related by the $\Delta_1\rightarrow e^{i\theta}\Delta_1$ and $\Delta_2\rightarrow e^{-i\theta}\Delta_1$ transformation.
This continuous degeneracy reflects the symmetry of Eq.(\ref{mf2}) under this  transformation. Under such circumstance there will be a Goldstone mode. To determine the effective action for the latter,
we let $\Delta_1(x,t)=\Delta_0e^{i\theta(x,t)}$ and 
$\Delta_2(x,t)=-\Delta_0e^{-i\theta(x,t)}$ in the mean-field Hamiltonian and integrate out the electronic degrees of freedom \cite{nagaosa}. (In the above $\theta(x,t)$ is a smooth space-time function). To determine whether or not this mode is charged, we impose a background electric field $E$. The details of such calculation will be reported elsewhere \cite{dhl}, but the following is the answer. After proper rescaling of $x$ and $t$ the effective action for $\phi$ is 
\eq
S_g=\frac{K}{2}\int dtdx (\phi^*\frac{\p_{\mu}}{i}\phi)^2,
\label{gauss}
\ee 
where $\phi\equiv e^{i\theta}$.
The lack of electric field dependence in Eq.(\ref{gauss}) indicates that the Goldstone mode is neutral.
 
{\bf{II. The doped case.}}
\rm
In the doped case $g_4^3=g_1^3=g_2^3=0$. Integrating Eq.(\ref{rg}) numerically for $V/U<1$, we see that the absolute values of all coupling constants again  eventually diverge. In this case there are three phases. For  $\frac{V}{U}<(\frac{V}{U})_{c1}\approx 0.55$ the 
most divergent susceptibility is the superconducting $\chi_{SS2}$! For $(\frac{V}{U})_{c1}<\frac{V}{U}<(\frac{V}{U})_{c2}\approx 0.65$ the most divergent susceptibility is $\chi_{SDW1}$. Finally for $(\frac{V}{U})_{c2}<\frac{V}{U}<1$
the most divergent susceptibility is $\chi_{CDW1}$.

{\bf{IIa) The SS2 phase}}
\rm
For $\frac{V}{U}<(\frac{V}{U})_{c1}$ we introduce
$\Delta_1=<b_1(x)>$ and $\Delta_2=<b_2(x)>$ as the order parameters where 
\eqa
&&b_1^+(x)=\frac{1}{\sqrt{2}}[\psi_{R+\uparrow}^\dagger(x) \psi_{L-\downarrow}^\dagger(x)- \psi_{R+\downarrow}^\dagger(x)\psi_{L-\uparrow}^\dagger(x)]\nonumber \\
&&b_2^+(x)=\frac{1}{\sqrt{2}}[\psi_{R-\uparrow}^\dagger(x) \psi_{L+\downarrow}^\dagger(x)- \psi_{R-\downarrow}^\dagger(x)\psi_{L+\uparrow}^\dagger(x)].
\eea
The mean-field Hamiltonian is the factorized form of \cite{tb},
\eqa
H&&=H_K+(g_1^1+g_2^2)\int dx[b_1^+(x)b_1(x)+b_2^+(x)b_2(x)\nonumber \\
&&+(g_1^2+g_2^1)\int dx[b^+_1(x)b_2(x)+b^+_2(x)b_1(x)].
\label{mf3}
\eea
For the range of $V/U$ considered here, after sufficient steps of renormalization $g_1^1+g_2^2$ becomes negative.
Meanwhile, $g_1^2+g_2^1$ stays positive. In the mean-field theory the term proportional to $(g_1^2+g_2^1)$ sets  $\Delta_1=-\Delta_2$. There is also a continuous family of mean-field solutions generated by the $\Delta_{1,2}\rightarrow e^{i\phi}\Delta_{1,2}$ transformation. This degeneracy reflects the global U(1) invariance of Eq.(\ref{mf3}).
The Goldstone mode associated with $\phi$  dominates the low energy physics of the SS2 phase.
If we let $\Delta(x,t)=\Delta_0e^{i\theta(x,t)}$, after proper rescaling of $x$ and $t$ the effective action for the Goldstone mode is 
\eq
S_g'=\frac{K}{2}\int dtdx [\phi^*(\frac{\p_{\mu}}{i}-2A_{\mu})\phi]^2.
\label{gauss2}
\ee
Here $\phi\equiv e^{i\theta}$ and $A_{\mu}$ is the external gauge field. The factor of two in front of $A_{\mu}$ reflects the fact that the Cooper pair is doubly charged.

Now we address the effect of impurities on superconductivity. 
Nanotubes often have impurities. For a single impurity the following terms are added to the Hamiltonian:
\eq
H_{imp}=\sum_{i,j=R,L}\sum_{a,b=\pm}u_{ij}^{ab}[\psi^+_{ia\s}(0)\psi_{jb\s}(0)+ h.c].
\ee
In the above ``0'' is the position of the impurity.
The one-loop RG equations for $u_{ij}^{ab}$ are:
\eqa
&&\frac{du_{LR}^{-+}}{dx}=\frac{1}{2}( g_2^2-2g_1^1) u_{LR}^{-+}\nonumber \\
&&\frac{du_{RR}^{-+}}{dx}=\frac{1}{2}(g_4^2+g_1^2-2g_4^1-2g_2^1)u_{RR}^{-+},
\eea
plus six other equations obtained by $R\leftrightarrow L$ and $+\leftrightarrow -$. All other $u's$ do not renormalize.
For positive bare $u_{LR}^{-+}$ and $u_{RR}^{-+}$, $u_{LR}^{-+}$ eventually grows upon renormalization, while $u_{RR}^{-+}$ eventually shrinks.  Nominally, we would drop the latter and keep only the former. However, within the range of coupling constants where we can trust our perturbative analyses, $u_{LR}^{-+}$ ($u_{RR}^{-+}$) is only amplified (suppressed) by roughly a factor of two. For this reason, in the following we shall analyze the effects of all impurity scattering amplitudes on Cooper pairs. 

We let an initial Cooper pair state $B^+(0)|0>\equiv\frac{1}{\sqrt{2}}(b_1^+(0)-b_2^+(0))|0>$
be scattered by vaious impurity scattering channels. The following are the results.
1) The terms introduced by $u_{LR}^{++},u_{LR}^{--},u_{RL}^{++},u_{RL}^{--}$ annihilate the pair. 2) The terms introduced by
$u_{RR}^{-+},u_{RR}^{+-},u_{LL}^{-+},u_{LL}^{+-},
u_{LR}^{-+},u_{LR}^{+-},u_{RL}^{-+},u_{RL}^{+-}$ break the pair. 3)
The terms introduced by $u_{RR}^{++},u_{RR}^{--},u_{LL}^{++},u_{LL}^{--}$ scatter it.
Therefore
a large amount of impurities can destroy superconductivity through 2), and produce Copper pair localization
through 3).

{\bf IIb) The SDW1 phase}
\rm
For $(\frac{V}{U})_{c1}<\frac{V}{U}<(\frac{V}{U})_{c2}$ our order parameters are the same as those defined in Eq.(\ref{order}). The mean-field Hamiltonian is the factorized form of Eq.(\ref{mf1}) except that $g_1^3$ and $g_4^3$ are set to zero. For the range of $V/U$ considered here the renormalized values of $g_4^2$ and $g_1^2$ are both positive. Consequently, the term proportional to $g_1^2$ requires $M_R=M_L$. In this case, in addition to the global SU(2) freedom associated with the spin rotation, there remains a global U(1) freedom, i.e., $M_{R,L}\rightarrow e^{i\phi}M_{R,L}$. 
The effective action for the spin Goldstone mode is the same as that in Eq.(\ref{sigma}), hence the spin excitations remain gapped. The effective action for the charge Goldstone mode is that of Eq.(\ref{gauss}) excepts that an additional term arising from the chiral anomaly should be added.
\eq
S_g''=S_g+S_{chiral};\hspace{0.2in}S_{chiral}=i\frac{4}{\pi}\int dxdt E\theta(x,t)
\label{gauss3}
\ee
The last term reflects the fact that the U(1) Goldstone mode is a charged mode. The physical meaning of this mode is the sliding degree of freedom of the spin density wave.

{\bf IIc) The CDW1 phase}
\rm
For $(\frac{V}{U})_{c2}<\frac{V}{U}<1$ we introduce $\Delta_R=<O_R(x)>$ and $\Delta_L=<O_L(x)>$ as order parameters. In the above
\eqa 
&&O_R(x)=\sum_{\s}\psi_{R+\sigma}^\dagger(x)\psi_{R-\sigma}(x) \nonumber \\
&&O_L(x)=\sum_{\s}\psi_{L+\sigma}^\dagger(x)\psi_{L-\sigma}(x).
\eea
The mean-field Hamiltonian is the factorized form of
\eq
H=H_K+\frac{1}{2}(2g_2^1-g_1^2)\int dx[O_RO_L^++O_LO_R^+].
\ee
In the range of $V/U$ considered here $2g_2^1-g_1^2<0$.
Consequently, mean-field theory gives $\Delta_R=\Delta_L$ with a global U(1) freedom, namely $\Delta_{R,L}\rightarrow e^{i\theta}\Delta_{R,L}$. The Goldstone mode associated
with this U(1) freedom is again governed by an action of the form Eq.(\ref{gauss3}). In this case $\phi$ describes the sliding mode of the charge density wave.      

In this paper we use an on-site and nearest-neighbor interaction to model the screened Coulomb interaction. The effects of an unscreened Coulomb potential  remain to be studied. In addition, we have not yet studied the nature of various phase transitions as a function of $V/U$.
Finally, our recursion relation (Eq.(\ref{rg})) has been obtained previously in studying various version of two-band models\cite{ladder}.
The relation between our model and the latter can be established via the identification that $(R,+)\leftrightarrow (A,R),
(R,-)\leftrightarrow (B,L), (L,+)\leftrightarrow (B,R),
(L,-)\leftrightarrow (A,L)$, where $A$ and $B$ labels the two bands.

Acknowledgement: We thank Professors Steven Kivelson, Paul McEuen and Doctor Vincent Crespi for helpful discussions. SGL acknowleges support of NSF Grant No. DMR 95-20554 and US DOE contract No. DE-AC03-76SF00098.


\begin{references}
\bibitem{note} A word of caution is in order here. As $n$ gets large, $E_c$ continuously decrease. Indeed, the energy range in which our description remains valid shrinks to zero as $n\ra\infty$!
\bibitem{note2} Here we have used the common convention, i.e., 1= inter-branch back scattering, 2=inter-branch forward scattering, 3=umklapp scattering, and 4=intra-branch forward scattering, to describe the structure among the upper and lower indices.
\end{references}
\end{document}